\documentclass[aps,prd,nofootinbib,twocolumn,floatfix,superscriptaddress]{revtex4}
\usepackage{graphicx}
\usepackage{dcolumn}
\usepackage{times,mathptm}
\newcommand{\beq}{\begin{equation}}
\newcommand{\eeq}{\end{equation}}
\newcommand{\bea}{\begin{eqnarray}}
\newcommand{\eea}{\end{eqnarray}}

\renewcommand{\d}{\delta}
\renewcommand{\l}{\lambda}

\renewcommand{\b}{\beta}

\renewcommand{\k}{\kappa}

\newcommand{\m}{\mu}

\newcommand{\s}{\sigma}

\newcommand{\D}{{\cal D}}

\newcommand{\oh}{{\textstyle{\frac{1}{2}}}}

\newcommand{\oq}{{\textstyle{\frac{1}{4}}}}

\newcommand{\non}{\nonumber}

\newcommand{\rf}[1]{(\ref{#1})}
\newcommand{\ra}{\rightarrow}
\newcommand{\pa}{\partial}

\begin{document}

\title{Dimensional Reduction and the Yang-Mills Vacuum State in 2+1 Dimensions} 

\author{J. Greensite}

\affiliation{The Niels Bohr Institute, DK-2100 Copenhagen \O, Denmark}
\affiliation{Physics and Astronomy Dept., San Francisco State
University, San Francisco, CA~94132, USA}

\author{{\v S}. Olejn\'{\i}k}
\affiliation{Institute of Physics, Slovak Academy
of Sciences, SK--845 11 Bratislava, Slovakia}
 
\date{\today}
\begin{abstract}
    We propose an approximation to the ground state
of Yang-Mills theory, quantized in temporal gauge and 2+1 dimensions, which 
satisfies the Yang-Mills Schr{\"o}dinger equation in both the free-field 
limit, and in a strong-field zero mode limit. Our proposal contains a 
single parameter with dimensions of mass;  confinement via dimensional 
reduction is obtained if this parameter is non-zero, and a non-zero 
value appears to be energetically preferred.  A method for numerical
simulation of this vacuum state is developed.  It is shown that
if the mass parameter is fixed from the known string tension in
2+1 dimensions, the resulting mass gap deduced from the vacuum
state agrees, to within a few percent, with known results for the mass gap 
obtained by standard lattice Monte Carlo methods.  
\end{abstract}

\maketitle
%
%
 
\section{\label{sec:intro}Introduction}

       Confinement is a property of the vacuum state of quantized 
non-abelian gauge theories,  and it seems reasonable that something
could be learned about the origin of confinement, and the origin of the mass gap, 
if we knew the form of the Yang-Mills vacuum wavefunctional in some physical
gauge.  There have, in fact, been a number of efforts along those lines, 
in temporal gauge \cite{Me1,Me2,Marty,Feynman,Mansfield,Kovner,Samuel}, 
Coulomb gauge \cite{Adam,Hugo}, axial gauge \cite{Peter}, and in a Bars corner-variable
formulation \cite{KKN,Rob}.

    In this article we will pursue this investigation in temporal gauge and in
$D=2+1$ dimensions, our strongest influences being refs.\ \cite{Me1} and \cite{Samuel}.
Our claim is that the ground state wavefunctional $\Psi_{0}[A]$ can be approximated
by the form 
\bea
\Psi_0[A] &=& \exp\left[-\oh\int d^2x d^2y ~ B^a(x) \right.
\non \\
  & & \qquad \left.  \left({1 \over \sqrt{-D^2 - \l_0 + m^2}} 
\right)^{ab}_{xy} B^b(y) \right]      
\label{vacuum}
\eea 
where $B^{a}=F_{12}^{a}$ is the color magnetic field strength, $D^{2}=D_{k}D_{k}$ is the two-dimensional
covariant Laplacian in the adjoint color representation, $\l_{0}$ is the lowest eigenvalue of $-D^{2}$, and
$m$ is a constant, with dimensions of mass, proportional to $g^{2}$.  To support this claim, we will argue 
that the above expression
\begin{enumerate}
\item is the ground state solution of the Yang-Mills Schr{\"o}dinger equation in the $g\ra 0$ limit;
\item solves the zero-mode Yang-Mills Schr{\"o}dinger equation in the zero-mode strong-field limit;
\item confines if $m>0$, and that $m>0$ is energetically preferred;
\item results in the numerically correct relationship between the mass gap and string tension.
\end{enumerate} 
A very similar proposal for the vacuum wavefunctional, with $\l_{0}$ absent, was put forward by Samuel
in ref.\ \cite{Samuel}, generalizing the earlier ``dimensional reduction" proposal of ref.\ \cite{Me1}. 

    Our paper is organized as follows:  In section \ref{sec:zmode}, below, we find an approximate solution
of the zero mode Yang-Mills Schr{\"o}dinger equation in 2+1 dimensions, and compare this to our proposed wavefunctional 
in an appropriate limit.  The dimensional reduction and confinement properties are discussed 
in section \ref{sec:dimred}.  Section \ref{sec:numsim} outlines a procedure for numerical simulation of 
our vacuum wavefunctional; in section \ref{sec:massgap} this procedure is applied to calculate the mass gap, 
with parameter $m$ chosen to give the correct string tension as a function of coupling.  Confinement, 
in our approach, relies on $m^{2} > 0$; in section \ref{sec:vacuumenergy} we will discuss why this choice 
lowers the vacuum energy in the non-abelian theory, while the minimum is at $m^{2}=0$ in the free 
abelian theory.  Section \ref{sec:KKN} contains a few results and critical comments 
regarding certain other proposals for the Yang-Mills vacuum wavefunctional. 
Some brief remarks about Casimir scaling and N-ality are 
found in section \ref{sec:nality}, with conclusions in section \ref{sec:conclusions}.

   We would like to note here that the work in section \ref{sec:zmode}, concerning the zero-mode strong-field limit,  
was motivated by a private communication from D.\ Diakonov to one of the authors \cite{Mitya}.

\section{\label{sec:zmode}The Free Field and Zero Mode Limits}
        
        In temporal gauge and $D=d+1$ dimensions, the problem is to find the ground state of the Yang-Mills
Schr{\"o}dinger equation
\beq
      H \Psi_0 = E_0 \Psi_0
\eeq
where
\beq
H = \int d^dx \left\{ -\oh {\d^2 \over \d A^a_k(x)^2} + \oq 
F_{ij}^{a}(x)^2 \right\}
\eeq
and all states in temporal gauge, in SU(2) gauge theory, are subject to the physical state condition
\beq
\Bigl( \d^{ac} \pa_k + g \epsilon^{abc} A^b_k \Bigr) {\d \over \d A^c_k }\Psi = 0 
\eeq
This condition requires invariance of $\Psi[A]$ under infinitesimal gauge transformations.

     Our proposed vacuum wavefunctional, eq.\ \rf{vacuum}, obviously satisfies the physical state
condition, since the kernel
\beq
           K_{xy}^{ab} = \left( {1\over \sqrt{-D^2 - \l_0  + m^2} }\right)^{ab}_{xy}
\eeq
transforms bilinearly, $K_{xy}\ra U(x)K_{xy}U^{-1}(y)$, under a gauge transformation, with $U$ a transformation
matrix in the adjoint representation.    In the $g\ra 0$ limit, with both $\l_{0},m \ra 0$ in the same limit,
the vacuum state becomes
\bea
\Bigl(\Psi_0[A]\Bigr)_{g\ra 0} 
 &=& \exp\left[-\oh \int d^2x d^2y \Bigl(\pa_1 A^a_2(x) - \pa_2 A^a_1(x) \Bigr) \right.
\non \\
          & & \left. \left({\d^{ab} \over \sqrt{-\nabla^2}} \right)_{xy} 
\Bigl(\pa_1 A^b_2(y) - \pa_2 A^b_1(y) \Bigr) \right]
\label{abelian-vac}
\eea
which is the known ground state solution in 2+1 dimensions, in the abelian, free-field case.  

    The Yang-Mills Schr{\"o}dinger equation is also tractable in a quite different limit, which is, in a sense,
diametrically opposed to the free-field situation.  Let us restrict our attention to gauge fields which are constant
in the two space directions, and vary only in time (analogous to the minisuperspace approximation
in quantum gravity).  The Lagrangian is
\bea
      L &=& \oh \int d^{2}x ~ \Bigl[ \pa_{t} A_{k}\cdot \pa_{t} A_{k}
             - g^{2} (A_{1} \times A_{2}) \cdot (A_{1} \times A_{2}) \Bigr]
\non \\
      &=&  \oh V \Bigl[ \pa_{t} A_{k}\cdot \pa_{t} A_{k}
             - g^{2} (A_{1} \times A_{2}) \cdot (A_{1} \times A_{2}) \Bigr]     
\eea
where $V$ is the area of a time-slice, leading to the Hamiltonian operator
\beq
     H = -\oh {1\over V} {\pa^{2} \over \pa A_{k}^{a} \pa A_{k}^{a} }
                 + \oh g^{2} V  (A_{1} \times A_{2}) \cdot (A_{1} \times A_{2}) 
\eeq 

    The factors of $V$ in the Hamiltonian suggest the use of a $1/V$
expansion.  Let us write
\beq
           \Psi_{0} = \exp[-VR_{0} + R_{1} + V^{-1}R_{2} +...)]
\label{WKB}
\eeq
with $R_{0}$ chosen such that the leading order (in $1/V$) ``kinetic" term contained
in $H\Psi_{0}$
\beq
       -\oh V {\pa R_{0} \over \pa A_{k}^{a}}{\pa R_{0} \over \pa A_{k}^{a}} \Psi_{0}
\eeq
cancels the potential term
\beq
    \oh g^{2} V (A_{1} \times A_{2}) \cdot (A_{1} \times A_{2}) \Psi_{0}
\eeq
at $O(V)$.  Let
\beq
            R_{0} = \oh g  {(A_{1} \times A_{2}) \cdot 
                      (A_{1} \times A_{2}) \over \sqrt{|A_{1}|^{2} + |A_{2}|^{2}}}
\label{R0}
\eeq
Then, defining
\beq
     T_{0} =   V\left[ -{\pa R_{0} \over \pa A_{k}^{a}}{\pa R_{0} \over \pa A_{k}^{a}}
             + g^{2} (A_{1} \times A_{2}) \cdot (A_{1} \times A_{2}) \right] 
\eeq
it is not hard to verify that
\bea
    T_{0} &=& 0 +  {7\over 4} g^{2} V {\Bigl[ (A_{1} \times A_{2}) \cdot (A_{1} \times A_{2})\Bigr]^{2}
                      \over \Bigl(|A_{1}|^{2} + |A_{2}|^{2}\Bigr)^{2}} 
\non \\
            &=& { 7 V R_{0}^{2} \over |A_{1}|^{2} + |A_{2}|^{2}}
\label{T0}
\eea
Now for $A$-fields for which $\Psi_{0}$ is non-negligible, 
it is easy to see that $T_{0} \Psi_{0}$ is of order no greater than $1/V$, except in the immediate
neighborhood of the origin ($A_{k}=0$) of field space.  That is because
$\Psi_{0}\approx \exp[-V R_{0}]$, which is non-negligible only if $VR_{0}$ is $O(1)$. 
For comparison with eq.\ \rf{vacuum} we are interested in a strong-field limit, far from
the origin of field space.  In that case, since $R_{0} \sim 1/V$, then the rhs of \rf{T0}
is at most of order $1/V$, which can  be neglected.   It follows that $R_{0}$ 
in eq.\ \rf{R0} accomplishes the required cancellation at leading order, and provides 
the leading contribution to the logarithm of the vacuum wavefunction.  

        Now consider the proposal \rf{vacuum} for the vacuum wavefunctional of the full theory, in a
corner of field space where the non-zero momentum modes of the $A$-field are negligible compared
to the zero modes, and  in fact the zero modes are so large in magnitude that we can approximate
$D_{k}^{ac} \approx g\epsilon^{abc}A^{b}_{k}$.  In this region
\beq
           (-D^{2})^{ab}_{xy} = g^{2} \d^{2}(x-y) M^{ab}
\eeq
where
\beq
            M^{ab} = (A_{1}^{2}+A_{2}^{2})\d^{ab} - A_{1}^{a}A_{1}^{b} -  
                                A_{2}^{a}A_{2}^{b}     
\eeq
In SU(2) gauge theory, the two zero-mode fields $A_{1},~A_{2}$ define a plane in
three-dimensional color space.  Take this to be, e.g., the color $x-y$ plane, i.e.
\beq
         A_{1} = \left[\begin{array}{ccc}
                                a_{1} \cr
                                a_{2} \cr
                                   0 \end{array}\right]  ~~~,~~~ 
         A_{2} = \left[\begin{array}{ccc}
                                b_{1} \cr
                                b_{2} \cr
                                   0 \end{array}\right]  ~~~,~~~ 
\label{As}
\eeq
Then
\beq
      M = \left( \begin{array}{ccc}
                 a_{2}^{2} + b_{2}^{2} & - a_{1}a_{2}- b_{1}b_{2}  &   0     \cr
              - a_{1}a_{2}- b_{1}b_{2} &   a_{1}^{2} + b_{1}^{2}   &   0     \cr
                               0                &                  0                  &    A_{1}^{2}+A_{2}^{2} \end{array}  
               \right) 
\eeq
Now $M$ has three eigenstates
\beq
           \phi_{1}=\left[\begin{array}{ccc}
                             \phi_{1}^{1} \cr
                             \phi_{1}^{2} \cr
                                   0 \end{array}\right]  ~~~,~~~
           \phi_{2}=\left[\begin{array}{ccc}
                             \phi_{2}^{1} \cr
                             \phi_{2}^{2} \cr
                                   0  \end{array}\right]  ~~~,~~~
            \phi_{2}=\left[\begin{array}{ccc}
                            0 \cr
                            0 \cr
                            1  \end{array}\right]
\label{phi}
\eeq
with corresponding eigenvalues
\bea
       \m_{1} &=& \oh \left( S - \sqrt{S^{2}-4C} \right)
\non \\
       \m_{2} &=& \oh \left( S + \sqrt{S^{2}-4C} \right)
\non \\
        \m_{3} &=& S
\eea
where
\beq
          S = A_{1}^{2} + A_{2}^{2}  ~~~,~~~  C = (A_{1} \times A_{2})\cdot  (A_{1} \times A_{2})
\eeq           
Then
\beq
           \left({1\over \sqrt{M - (\m_{1} - m^{2}) I}}\right)^{ab} = 
                   \sum_{n=1}^{3} {\phi^{a}_{n} \phi^{*b}_{n}\over \sqrt{\m_{n} - \m_{1} + m^{2}}}
\label{sum}
\eeq
We have
\bea
 \Psi_{0} 
 &\approx& \exp\left[- \oh \int d^{2}x d^{2} y ~ B^{a}(x) 
                      \left({1\over\sqrt{-D^{2} - \l_{0} + m^{2}}}\right)^{ab}_{xy} \right.
\non \\
    & &     \left. \qquad B^{b}(y) \right]
\non \\
     &=&  \exp\left[- \oh g^{2} V (A_{1} \times A_{2})^{a}  \left({1\over \sqrt{g^{2}(M - \m_{1} I) + m^{2} I}}\right)^{ab} \right.
\non \\
     & &     \left. \qquad  (A_{1} \times A_{2})^{b}   \right]
\eea
Taking account of eqs.\ \rf{As}, \rf{phi} and \rf{sum}, we get
\bea
\Psi_{0} &=& \exp\left[- \oh g V (A_{1} \times A_{2})^{3}\left({1\over \sqrt{M - \m_{1} I + m^{2} I}}\right)^{33} \right.
\non \\
        & & \qquad           \left.   (A_{1} \times A_{2})^{3}   \right]
\non \\
         &=& \exp\left[- \oh g V {(A_{1} \times A_{2})\cdot
                      (A_{1} \times A_{2}) \over \sqrt{\m_{3} - \m_{1} + m^{2}}}    \right]
\label{psi1}
\eea
Now by assumption, in the strong-field limit,
\beq
          g^{2} \m_{3} = g^{2}(A_{1}^{2} + A_{2}^{2} ) \gg m^{2}
\eeq
and
\bea
           \m_{1} &=& \oh S \left(1 - \sqrt{1 - 4{C\over S^{2}}}\right)
              \approx  {C\over S} 
\non \\
                 &\approx& {2\over g} R_{0}
\eea
 We recall that the ground-state solution of the zero-mode Schr{\"o}dinger equation 
$\Psi_{0} = \exp[-VR_{0}]$ with $R_{0}$ given in eq.\ \rf{R0} is
valid for $R_{0} \sim 1/V$, where the wavefunction is non-negligible.  In this same
region of configuration space, $\m_{1}$ is negligible compared to $\m_{3}$, and eq.\
\rf{psi1} becomes
\bea
\Psi_{0}  &=&  \exp\left[- \oh g V {(A_{1} \times A_{2})\cdot
                      (A_{1} \times A_{2}) \over \sqrt{\m_{3}} } \right]
\non \\
   &=&  \exp\left[- \oh g V {(A_{1} \times A_{2})\cdot
                      (A_{1} \times A_{2}) \over \sqrt{A_{1}^{2} + A_{2}^{2}} } \right]
\eea
which is identical to the solution found for the ground state of the zero-mode Schr{\"o}dinger
equation, in the region of validity of that solution, where $VR_{0} \sim O(1)$ .
Therefore, in a small region of configuration space where a non-perturbative treatment
is possible, we find that our ansatz for the vacuum state agrees with the ground state
of the zero-mode Yang-Mills Schr{\"o}dinger equation.\footnote{We learned
from D.\ Diakonov that he had obtained this result in unpublished work, which considered
a wavefunctional of similar form to \rf{vacuum} but without the $\l_{0},m$ terms 
in the kernel \cite{Mitya}.  Those terms are not important in the region of configuration
space discussed in this section.}

     The argument above can also be extended to 3+1 dimensions, as outlined in Appendix \ref{sec:appA}.

\section{\label{sec:dimred}Dimensional Reduction and Confinement} 

    Assuming that our proposal \rf{vacuum} for the Yang-Mills vacuum wavefunctional
in 2+1 dimensions is at least approximately correct, then where does the confinement
property appear?

    A long time ago it was suggested that the effective Yang-Mills vacuum wavefunctional at large
scales, in $D=d+1$ dimensions, has the form \cite{Me1}
\beq
           \Psi_0^{eff} \approx \exp\left[ -\mu \int d^dx ~ F^a_{ij}(x) F^a_{ij}(x) \right]
\label{dimred}
\eeq
(see also \cite{Marty,Mansfield}).
This vacuum state has the property of ``dimensional reduction":  Computation of a 
large spacelike loop in $d+1$ dimensions reduces to the calculation of a large Wilson 
loop in $d$ Euclidean dimensions.  Suppose  $\Psi_{0}^{(3)}$  is the ground state of the 3+1 
dimensional theory, and $\Psi_{0}^{(2)}$  is the ground state of the 2+1 dimensional theory.   
If these ground states both have the dimensional reduction form, and W(C) is a large planar 
Wilson loop, then the area law falloff in $D=3+1$
dimensions follows from confinement in two Euclidean dimensions in two steps:
\bea
           W(C) &=& \langle \mbox{Tr}[U(C)]\rangle^{D=4} 
                      = \langle \Psi^{(3)}_{0}|\mbox{Tr}[U(C)]|\Psi_0^{(3)}\rangle
\non \\
                        &\sim&  \langle \mbox{Tr}[U(C)]\rangle^{D=3} 
                      = \langle \Psi^{(2)}_{0}|\mbox{Tr}[U(C)]|\Psi_0^{(2)}\rangle
\non \\
                        &\sim& \langle \mbox{Tr}[U(C)]\rangle^{D=2} 
\eea
In $D=2$ dimensions the Wilson loop can of course be calculated analytically, and we know 
there is an area-law falloff, with Casimir scaling of the string tensions.  The dimensional reduction
form of the ground state wavefunctional can be demonstrated explicitly in strong-coupling
lattice gauge theory \cite{Me2}; Monte Carlo support for the hypothesis has also been obtained
at intermediate couplings \cite{Me3,Arisue}.

    It is natural to try and improve on the dimensional reduction idea
by considering wavefunctionals which interpolate, in some natural way,
between free-field dynamics at short distance scales, and the
dimensional reduction form at large scales.  In ref.\ \cite{Samuel},
Samuel suggested that the vacuum state in $D=2+1$ dimensions might
have the form
\beq
\Psi_0[A] = \exp\left[-\oh\int d^2x d^2y ~ B^a(x) 
\left({1 \over \sqrt{-D^2 + m_0^2}} \right)^{ab}_{xy} B^b(y) \right]
\label{stuart}
\eeq
Our proposal differs from Samuel's in that $m_{0}^{2}$ is replaced by
$-\l_{0} + m^{2}$, with the lowest eigenvalue $\l_{0}$ being
field-dependent and gauge-invariant.  The rationale is that we should
allow for a subtraction in the operator $-D^2$ appearing in the vacuum
kernel; a subtraction will be absolutely required if the spectrum of
$-D^2$, starting with $\l_0$, diverges in the continuum limit.
On the other hand, if $m_{0}^{2} < 0$ is a negative constant, then the
wavefunctional in eq.\ \rf{stuart} is not necessarily real throughout
configuration space, and can oscillate.  Now the true vacuum state
must be real up to a constant factor, and it is forbidden to pass
through zero by the ``no node'' theorem for quantum-mechanical ground
states.  Requiring a subtraction which respects the reality of the
wavefunctional, and avoids oscillations anywhere in field
configuration space, dictates the replacement
\beq
             m_{0}^{2} \ra -\l_{0}  + m^{2}
\eeq
with $m^2 \ge 0$.

    The dimensional reduction form is obtained by dividing the field
strength into ``fast" and ``slow" components, defined in terms of a
mode cutoff.  Let $\{\phi_{n}^{a}\}$ and $\{\l_{n}\}$ denote the
eigenmodes and eigenvalues, respectively, of the covariant Laplacian
operator in adjoint color representation, i.e.
\beq
        -(D^{2})^{ab}\phi_{n}^{b} = \l_{n} \phi_{n}^{a}
\eeq
The field strength can be expanded as a mode sum
\beq
B^{a}(x) = \sum_{n=0}^{\infty} b_{n} \phi^{a}_{n}(x)                 
\label{bn}
\eeq
and we define the "slow" component to be
\beq
 B^{a,{\rm slow}}(x) = \sum_{n=0}^{n_{max}} b_{n} \phi^{a}_{n}(x)        
\label{Bslow}
\eeq
where $n_{max}$ is a mode cutoff chosen such that $\Delta \l \equiv \l_{n_{max}}-\l_{0} \ll m^{2}$
remains fixed as $V\ra \infty$. In that case, the portion of the
(squared) vacuum wavefunctional gaussian in $B^{slow}$ is approximately
\beq
     \exp\left[-{1\over m}\int d^2x ~ B^{a,slow} B^{a,slow}\right]
\label{dimred1}
\eeq
which is just the probability measure for Yang-Mills theory in two Euclidean
dimensions, with a particular type of ultraviolet cutoff. The string tension 
for fundamental representation Wilson loops
in $D=2$ Yang-Mills theory, with coupling $g^{2} m$, is easily computed:
\beq
          \s = {3\over 16} g^{2} m
\eeq
or in lattice units, with lattice coupling $\b$,
\beq
          \s = {3\over 4} {m\over \b}
\label{sm}
\eeq          
       
       In the next sections we will address two questions.  First, suppose we fix $m$
to give the known string tension at a given lattice coupling.  What is then the value of
the mass gap predicted by the vacuum wavefunctional, and to what extent does
this agree with the corresponding value determined by standard lattice Monte Carlo
methods?  Secondly, since confinement depends on having $m\ne 0$, is there any reason 
why the mass parameter $m$ should be non-zero?

\section{\label{sec:numsim}Numerical Simulation of the Vacuum Wavefunctional}

     The mass gap implied by the vacuum state \rf{vacuum} can, in principle, be
extracted from the equal-times connected correlator
\beq
       \D(x-y) = \langle (B^a B^a)_x (B^b B^b)_y \rangle - \langle (B^a B^a)_x \rangle^2
\label{concorr}
\eeq
where the expectation value is taken with respect to the probability distribution
$P[A]$ defined by the vacuum wavefunctional, i.e.
\beq
\langle Q \rangle = \int DA_{1} DA_{2} ~ Q[A] P[A]
\eeq
with
\bea
P[A] &=& |\Psi_0[A]|^2 
\non \\
&=& \exp\left[- {1\over g^2}\int d^2x d^2y ~ B^a(x) K^{ab}_{xy}[A] B^b(y) \right]
\label{g2}
\eea
and
\beq
 K_{xy}^{ab}[A] = \left({1 \over \sqrt{-D^2 -  \l_0   + m^2}} \right)^{ab}_{xy} 
\eeq
Here we have absorbed a factor of $g$ into the definition of $A_i$, which accounts for the
factor of $1/g^2$ in the exponent in eq.\ \rf{g2}. 

It not easy to see how $\D(x-y)$ could be computed analytically beyond the level of weak-coupling
perturbation theory, but computation by numerical simulation of $P[A]$ also seems hopeless, at
least at first sight.  Not only is the kernel $K_{xy}^{ab}$ non-local, it is not even known explicitly
for arbitrary $A_{k}^{a}(x)$.   However, suppose that after eliminating the wild variations of $K$ along 
gauge orbits via a gauge choice, $K[A]$ has very little variance among thermalized configurations.
In that case, things are more promising.  

    Let us define a probability distribution for gauge fields $A$ which is controlled by a second,
independent configuration $A'$ 
\bea
\lefteqn{P\Bigl[A;K[A']\Bigr] =} 
\non \\
& &  {\det}^{1/2}\left[{1\over g^2}K[A']\right]
\exp\left[-{1\over g^2}\int d^2x d^2y ~ B^a(x) K^{ab}_{xy}[A'] B^b(y) \right]
\non \\
\label{PAK}
\eea
where the field strength $B$ is computed from the $A$-configuration, and both $A$ and $A'$
are fixed to some appropriate gauge. Now, assuming that
the variance of $K[A]$ in the probability distribution
$P[A]$ is small after the gauge choice, we can approximate
\bea
P[A]  &\approx& P\Bigl[A,\langle K \rangle \Bigr]
\nonumber \\
&=& P\left[A, \int DA' ~ K[A'] P[A'] \right]
\nonumber \\
&\approx& \int DA' ~ P\Bigl[A,K[A']\Bigr] P[A']
\label{PA}
\eea
where the step from the second to the third line follows from assuming that the variance 
of $K$ in the distribution $P[A]$ is small.   If this assumption about $K[A]$ is correct,
and eq.\ \rf{PA} holds, then the probability distribution could in principle be generated by
solving \rf{PA} iteratively:
\bea
P^{(1)}[A] &=& P\Bigl[A;K[0]\Bigr]
\non \\
P^{(n+1)}[A] &=& \int DA' ~ P\Bigl[A;K[A']\Bigr] P^{(n)}[A']
\eea    
A numerical version of this approach would be to use equibilibrium configurations of
$P^{(n)}[A]$, generated at the $n$-th step, to generate equilibrium configurations of 
$P^{(n+1)}$ at the $(n+1)$-th step.    

      We may use the remaining gauge freedom to fix to an axial gauge in the
two-dimensional time slice.   This allows us to change variables in the functional
integral over two-dimension configurations from $A_{1}^{a},~A_{2}^{a}$ to $B^{a}$, without 
introducing a field-dependent Jacobian.   Let eigenvalues $\l_{n}$, and eigenmodes 
$\phi_{n}^{a}$ solve the eigenvalue equation 
\beq
          -D^2 \phi_n = \l_n \phi_n
\eeq
for the covariant Laplacian $-D^{2}$ determined from the fixed $A'$ configuration, and let 
$\{b_{n}\}$ be the mode amplitudes of the B-field, as seen in the mode expansion \rf{bn}.
Then the probability distribution for the $\{b_{n}\}$, which follows from $P[A;K[A']]$
at fixed $A'$, is Gaussian
\beq
        \mbox{prob}[b_n] \propto \exp\left[-{\b \over 4} {b_n^2 \over \sqrt{\l_n - \l_{0} + m^{2}}} \right]
\eeq

    In practice we use a lattice regularization on an $L\times L$ lattice with periodic
boundary conditions, and the gauge field $A^{a}_{k}(x,y)$ is initialized to zero at the
first iteration.   We then generate gauge fields
recursively; the procedure at the $n$-th iteration is as follows:
\begin{enumerate}
\item From one of the lattice configurations generated at the $(n-1)$-th iteration,
compute the link variables in the adjoint representation, and then 
determine numerically the eigenvalues and eigenmodes of the two-dimensional lattice 
covariant Laplacian operator $-D^{2}$.
\item Generate a set of $3L^2$ normally-distributed 
random numbers with unit variance, denoted $\{r_n\}$.  From these, we obtain
a new set of mode amplitudes
\beq
        b_n = \sqrt{{2\over \beta}}(\l_n - \l_{0} + m^{2})^{1/4} r_n
\eeq
and a corresponding $B$-field
\beq
B^{a}(x) = \sum_{n=0}^{3L^{2}} b_{n} \phi^{a}_{n}(x)                 
\eeq
From the field strength $B^{a}(x)$, and the axial gauge condition, determine the corresponding 
gauge field $A_{k}^{a}(x)$.  This step can be repeated to generate as many thermalized 
configurations of $P[A,K[A']]$ as desired.
\item The gauge fields are exponentiated to give link variables
\beq
         U_k(x,y) = \exp[i A_k^a(x,y) \s_a /2]
\eeq
and any observables of interest are computed.  This concludes
the $n$-th iteration.
\end{enumerate}
Lattice configurations generated by this procedure will be referred to as
``\emph{recursion lattices}".

   Details about our particular choice of axial gauge on a finite lattice, and the procedure for
obtaining the $A$-field from the $B$-field in that gauge, may be found in 
Appendix \ref{sec:appB}.

\section{\label{sec:massgap}The Mass Gap}

    The simulation procedure outlined in the last section leans heavily on the assumption
that there is little variance in the kernel $K^{ab}_{xy}$ in a fixed gauge, or, equivalently,
that there is negligible variance, among thermalized configurations, 
in gauge-invariant combinations of the kernel such as Tr$[K^{-1}_{xy}K^{-1}_{yx}]$, or in the 
gauge-invariant spectrum of $K$.  The absence of significant fluctuations in these 
quantities, when evaluated numerically, is a self-consistency requirement of the method we
have proposed.   The quantity  Tr$[K^{-1}_{xy}K^{-1}_{yx}]$ is of particular interest, because its
rate of falloff at large $|x-y|$ is determined by the mass gap.

\begin{figure}[htb]
\centerline{\scalebox{0.47}{\includegraphics{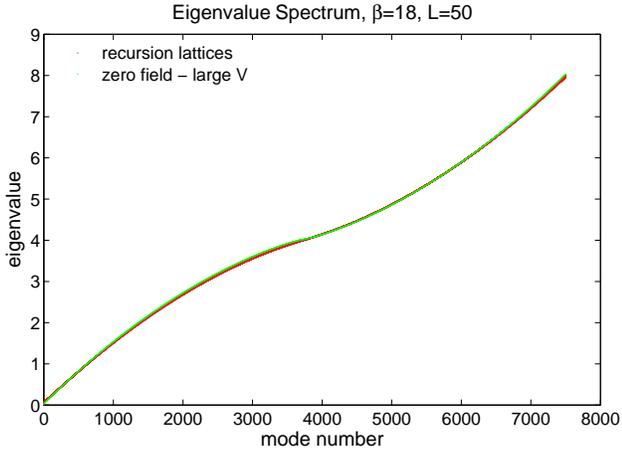}}}
\caption{Ten sets of eigenvalue spectra of the operator $-D^{2} -\l_{0}+m^{2}$, at $\b=18$, from ten independent
$50\times 50$ recursion lattices.   Also plotted, but indistinguishable from the other spectra, is the 
rescaled spectrum of the large-volume zero-field operator $-\nabla^{2} + m^{2}$. 
\label{eig}} 
\end{figure}

\begin{figure}[htb]
\centerline{\scalebox{0.47}{\includegraphics{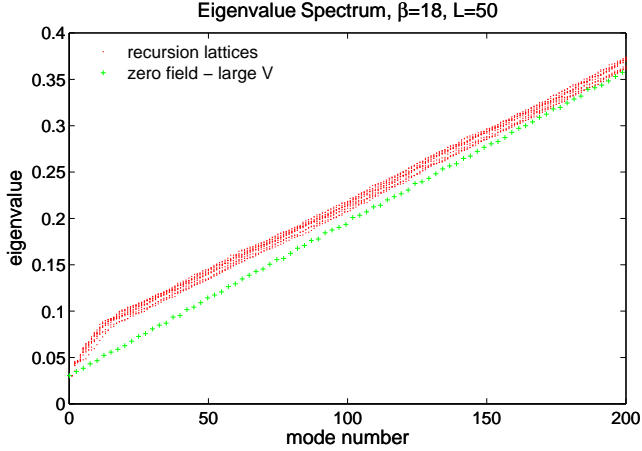}}}
\caption{Same as Fig.\ \ref{eig}, for the lowest 200 eigenmodes. The closely spaced dots are from ten sets of
eigenvalue spectra.  The ``+" symbols are taken from the rescaled spectrum of the large-volume zero-field
operator.
\label{eig1}} 
\end{figure}

    We begin with the spectrum $\{\l_{n}-\l_{0}+m^{2}\}$ of the operator $-D^{2} -\l_{0} + m^{2}$,
with $m$ chosen, at a given $\b$, to reproduce the string tension $\s(\b)$ known from Monte Carlo
simulations of the standard Wilson action in three Euclidean dimensions \cite{Teper}.  
From eq.\ \rf{sm}, this means choosing
\beq
            m = {4\over 3} \b \s(\b)
\label{mbs}
\eeq
The result for the spectrum at $\b=18$ on a $50\times 50$ lattice is
shown in Fig.\ \ref{eig}.  The figure displays our results for ten
separate recursion lattices, as well as the zero-field result
$-\nabla^{2} + m^{2}$ for a very large volume lattice, with the
eigenmode numbers rescaled by the factor $50^{2}/V$, so as to fit in
the same range on the x-axis as the other ten data sets.  It can be
seen that, at the resolution of this figure, the spectra essentially
all fall on top of one other.  The ten separate data sets cannot be
distinguished, and the spectrum of $-D^{2}-\l_{0}$ looks identical to
the (suitably rescaled) spectrum of $-\nabla^{2}$ at large volume.  At
higher resolution (Fig.\ \ref{eig1}) some fluctuation in the
eigenvalue spectrum is observable, and the eigenvalues of the
lowest-lying modes appear to deviate slightly from the zero-field
large-volume spectrum.

    Next we turn to the computation of the mass gap.  According to eq.\ \rf{PA},  
\bea
  \langle Q \rangle  &=& \int DA_{1} DA_{2} ~ Q[A] P[A]  
\non \\
 &\approx& \int DA_{1} DA_{2} DA'_{1} DA'_{2} ~ Q[A] P[A,K[A']] P[A']
\non \\
&=& \int DB  DA'_{1} DA'_{2} ~ Q[A(B)] P[A(B),K[A']] P[A']
\eea
where we have changed variables, in an axial gauge, from gauge field
$A$ to field strength $B$ as discussed in the last section.
Evaluating in this way the rhs of \rf{concorr} with $P[A,K[A']]$ as
defined by eq.\ \rf{PAK}, the integration over $B$ is gaussian, and we
find
\beq
\D(R) =  {8\over \beta^2} G(R)
\eeq
where $R=|x-y|$ and (no sum over $x,y$)
\bea
    G(R) &=&\Bigl\langle (K^{-1})^{ab}_{xy}(K^{-1})^{ba}_{yx} \Bigr\rangle
\non \\
K^{-1} &=& \sqrt{-D^{2} - \l_{0} + m^{2}}
\eea

    Of course, the  expectation value of $(K^{-1})^{ab}_{xy}(K^{-1})^{ba}_{yx}$  
can also be evaluated
by standard lattice Monte Carlo methods based on the $D=3$ dimensional
Wilson action.  A number of thermalized lattices are generated by the usual
heat bath procedure, and $K^{-1}$ is evaluated on a two-dimensional constant-time
slice of each three-dimensional lattice.  The two-dimensional lattices generated
in this way will be referred to as ``MC lattices".  They can be thought of as having
been drawn from a probability weighting $P[U] = \Psi_{E,0}^{2}[U]$, where $\Psi_{E,0}[U]$
is the ground state of the transfer matrix of the $D=3$ dimensional Euclidean lattice
gauge theory.

\begin{figure}[htb]
\centerline{\scalebox{0.70}{\includegraphics{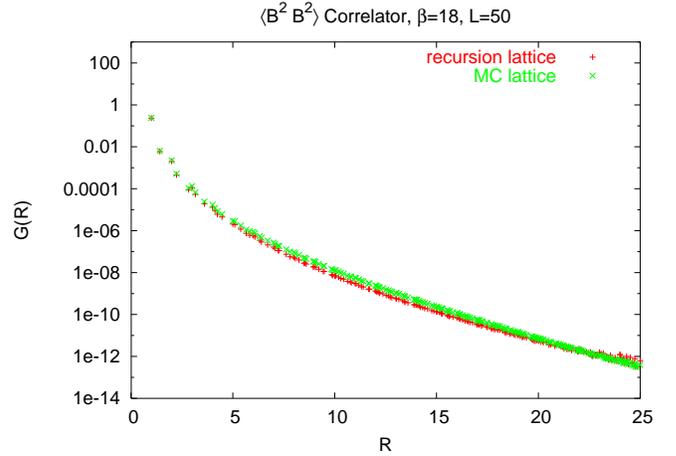}}}
\caption{The correlator $G(R)$ computed (i) on two-dimensional lattice configurations
generated from the vacuum wavefunctional by the method described in the text; and
(ii) on constant-time slices of three-dimensional lattice configurations generated by
the usual lattice Monte Carlo method. Lattices generated by the first method are denoted
``recursion", and by the second as ``MC".   In each case, the lattice extension is 50 sites
at $\b=18$.
\label{mass}} 
\end{figure}
    
     Figure \ref{mass}  shows the data for $G(R)$ at $\b=18$, averaged from a set of ten
$50\times 50$ recursion lattices,  and, for comparison, corresponding data averaged from a set 
of ten $50\times 50$ MC lattices at $\b=18$ .   
Note the very small ($\sim O(10^{-12})$) magnitude of the observable
at $R=20$, yet even at this magnitude there seems to be very little noisiness in the data.  
Once again, this absence of noise is only possible if the variance in the $K^{-1}K^{-1}$
observable is negligible, which supports our original hypothesis.  Moreover, the data obtained 
on recursion and MC lattices obviously agree very well with each other.

\begin{figure}[htb]
\centerline{\scalebox{0.70}{\includegraphics{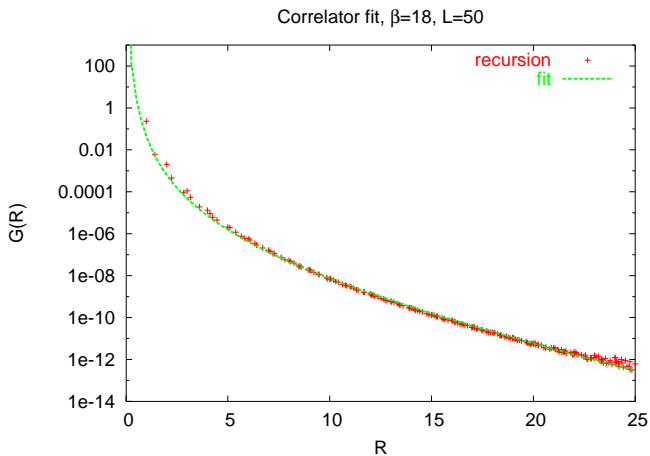}}}
\caption{Best fit (dashed line) of the recursion lattice data for $G(R)$ by the analytic
form given in eq.\ \rf{G0}.\label{fit}} 
\end{figure}

    The mass gap is obtained by fitting the data for $G(R)$ to an appropriate functional form, and
extracting the exponential falloff.  Define
\bea
G_{0}(R) &=& \d^{ab} \d^{ba}\left[ \left( \sqrt{-\nabla^2 + \m^2} \right)_{xy}\right]^2
\non \\
&=& {3\over 4\pi^2} (1 + \m R)^2 {e^{-2\m R} \over R^6}
\label{G0}
\eea
We have seen (Fig.\ \ref{eig}) that the spectrum of $-D^{2} -\l_{0}$ is almost identical
to that of the zero-field Laplacian $-\nabla^{2}$.  With this motivation, we introduce the fitting 
function
\beq
f_0(R) = \log\left[a (1 + \oh M R)^2 {e^{-M R} \over R^6} \right]
\label{Gfit}
\eeq
and carry out a two parameter ($a$ and $M$) best fit of $\log[G(R)]$ by $f_{0}(R)$.\footnote{The fits
were carried out by the GNUPLOT package, which implements the Marquardt-Levenberg fitting algorithm.
We have fit the data for $\log(G(R))$ on an $L\times L$ lattice in the interval $R\in [1,L/2]$.  Errorbars are
estimated from the variance in mass gaps computed separately, at each $\b$, on ten independent
lattices.}  The resulting value for $M$ is an estimate of the mass gap.
The best fit of the data for $G(R)$ at $\b=18$ on a $50^{2}$ lattice by the fitting function $\exp[f_{0}(R)]$ 
is shown in Fig.\ \ref{fit}. 
 
   In an old paper which anticipates the work in this section, Samuel \cite{Samuel} argued that $M\approx 2m_{0}$, where
$m_{0}$ is the mass parameter in the vacuum state \rf{stuart} which he proposed.  This result is obtained
if the covariant operator $-D^{2}$ in \rf{stuart} is replaced by $-\nabla^{2}$.   We believe that a more natural approximation is
the replacement of $-D^{2} -\l_{0}$ by $-\nabla^{2}$, since the lowest eigenvalue in the spectrum of each operator
begins at zero.  Thus the ``naive" estimate for the mass gap, in our proposal, is $M=2m$. 

\begin{figure}[t]
\centerline{\scalebox{0.70}{\includegraphics{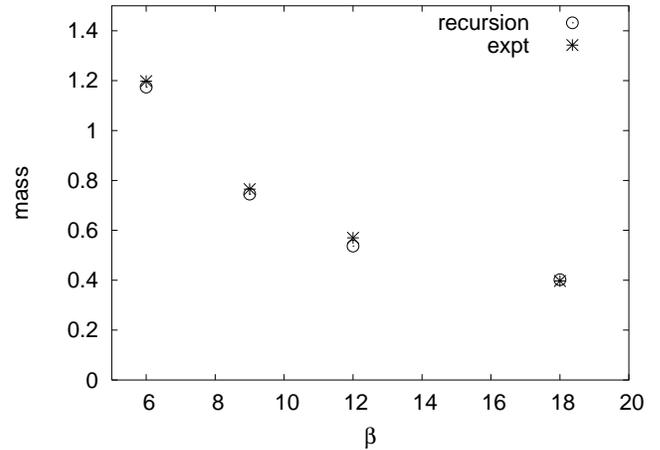}}}
\caption{Mass gaps extracted from recursion lattices at various lattice
couplings, compared to the $0^{+}$ glueball masses in 2+1 dimensions
obtained in ref.\ \cite{Teper} (denoted ``expt") via standard
lattice Monte Carlo methods. Errorbars are smaller than the symbol sizes.\label{gap}} 
\end{figure}

The results of extracting $M$ via the best fit of $f_{0}$ to the data,
for simulations of the vacuum wavefunctional at a variety of lattice
couplings, are shown in Fig.\ \ref{gap}.  There we compare our values
for the mass gap with those reported by Meyer and Teper in ref.\
\cite{Teper} (the values for $\s(\b)$, used in eq.\ \rf{mbs}, were
also taken from this reference.)  In Table \ref{table1} we list these
mass gap results, as well as the mass gaps extracted from MC lattices,
and the ``naive" estimate $M(0^{+})=2m$.  It can be seen that the
agreement between the reported values for the mass gap, and the masses
we have obtained from simulation of our proposed wavefunctional (with
parameter $m$ fixed to give the observed asymptotic string tension),
agree within a few ($<6$) percent. This is a substantial improvement
over the ``naive" estimate of $M=2m$, which disagrees with the Monte
Carlo results by up to $20\%$.

\begin{table}[htb]
\begin{center}
\begin{tabular}{|c|c|c|c|c|c|} \hline
 \multicolumn{2}{|c|}{ }&  \multicolumn{4}{c|}{mass gap} \\ \hline 
$\beta$ & $L^{2}$      & ``naive"       &  MC            & recursion   &   ``expt"          \\ 
             &                   &   ($M=2m$)  &  lattices      & lattices       &    ref. \cite{Teper} \\ \hline
     6  &       $24^{2}$  &   1.031         &  1.269(5)    &  1.174(8)    &  1.198(25)        \\ \hline
     9  &       $24^{2}$  &   0.627         &  0.775(3)    &  0.745(5)    &  0.765(8)          \\ \hline
    12 &       $32^{2}$  &   0.445         &  0.562(5)    &  0.537(5)    &  0.570(11)      \\ \hline
    18 &       $50^{2}$  &   0.349         &  0.436(3)     &  0.402(4)    &  0.397(8)         \\ \hline
\end{tabular}
\end{center}
\caption{The mass gaps in D=2+1 dimensional Yang-Mills theory, at a variety of $\b$ values and
lattice sizes $L^{2}$.  Column 3 shows the values derived from the estimate $M=2m$, and the values
extracted from $G(R)$ computed on MC lattices are shown in column 4.  Column 5 displays the
results extracted from $G(R)$ computed from recursion lattices; these are the predictions obtained
from numerical simulation of the vacuum wavefunctional.  All of these values can be compared to the
mass gaps reported in ref.\ \cite{Teper}, shown in column 6, which were
obtained by conventional lattice Monte Carlo methods.}
\label{table1}
\end{table}  
 
\section{\label{sec:vacuumenergy}Vacuum Energy and Confinement}      

    Our proposed vacuum wavefunctional results in a non-vanishing asymptotic string tension,
via the dimensional reduction argument, for any mass parameter $m>0$.  In this context, the
question of why pure SU(2) gauge theory confines in 2+1 dimensions boils down to why
$m$ is non-zero in that case, yet $m=0$ in the abelian theory.  The answer must lie in energetics:
For some reason the expectation value of $\langle H \rangle$ is lowered, in the non-abelian
theory, by having $m>0$.  

      The calculation of $\langle H \rangle$ is complicated by functional derivatives of the 
kernel $K[A]$.   In this initial study we will simply ignore these derivatives, on the grounds that
variance of the gauge-invariant product $K^{-1} K^{-1}$ among thermalized configurations has been
found, in numerical simulations, to be negligible.  In fact, this product seems to be remarkably well 
approximated, in any thermalized configuration, by the free-field expression $G_{0}(R)$ of eq.\ \rf{G0}.
This insensitivity to the $A$-field suggests that the variation of $K[A]$ in the neighborhood of 
thermalized configurations is extremely small, and therefore the neglect of functional derivatives of $K$ 
might be justified.  However, we are not as yet able to quantify the actual error which is made by
dropping those derivatives.

      Writing $\Psi_{0} = \exp(-R[A])$ where
\beq
           R = -{1\over 2 g^2} \int d^2x d^2y ~ B^a(x) K^{ab}_{xy} B^b(y)
\eeq
we find
\beq
       H \Psi_0 = \left( T_0 - T_1 + {1\over 2 g^2} \int d^2x ~ B^2 \right) \Psi_0
\eeq
where
\bea
          T_0 &=& {g^2\over 2} \int d^2 x ~ {\d^2 R \over \d A_k^c(x)^2}
\non \\
          T_1 &=& {g^2\over 2} \int d^2x ~ {\d R \over \d A_k^c(x)} {\d R \over \d A_k^c(x)}
\eea
Carrying out the indicated functional derivatives of $R$, but dropping terms involving
functional derivatives of the kernel $K$ leads to
\bea
         T_0 &=& \oh \int d^2x d^2y ~ \d(x-y) (-D^2)^{ab} K_{xy}^{ba}
\non \\
             &=& \oh \mbox{Tr}\left[ (-D^2) {1\over \sqrt{-D^2 - \l_{0} + m^2}} \right]
\eea
and
\bea
       T_1 &=& {1\over 2 g^2} \int d^2x d^2y ~ B^a(x) 
\left[ {-D^2 \over -D^2 -\l_{0}+ m^2} \right]_{xy}^{ab} B^b(y)
\non \\   
           &=& {1\over 2 g^2} \int d^2x d^2y ~ B^a(x) 
\left[ 1 + { \l_{0} - m^2 \over -D^2 - \l_{0} + m^2} \right]_{xy}^{ab} B^b(y)
\non \\
\eea
Altogether
\bea
 \langle H \rangle &=& 
       \oh \left\langle  \mbox{Tr}\left[ {-D^2 \over \sqrt{-D^2 - \l_{0} + m^2}} \right]  \right.
        - (\l_{0}-m^{2}) \times
 \non \\
    & &\left. {1\over g^2}\int d^2x d^2y ~ B^a(x)  
         \left( {1 \over -D^{2} -\l_{0} + m^{2}}\right)^{ab}_{xy} B^b(y) \right\rangle
 \non \\
\label{H1}
\eea

    Expanding $B(x)$ in eigenstates of $-D^2$
\beq
     B^a(x) = \sum_n b_n \phi^a_n(x)
\eeq
the second term on the rhs of eq.\ \rf{H1} becomes
\beq
   \mbox{2nd term} =  (\l_{0}-m^{2}){1\over g^2} \sum_n b_n^2 {1 \over \l_n - \l_0 + m^2}
\eeq
In the previous section, it was found that the eigenvalue spectrum $\{\l_n\}$ is almost unchanged from
one equilibrium lattice to the next.  Then, in the VEV shown in \rf{H1}, we may replace
$b_n^2$ by its VEV with the $\{\l_n\}$ fixed, which is $\oh g^2 \sqrt{\l_n - \l_0 + m^2}$.  Then
\bea
  \mbox{2nd term} &=&  \oh (\l_{0}-m^{2}) \sum_n {1\over \sqrt{\l_n - \l_{0} + m^2}}    
\non \\
          &=& \oh (\l_{0}-m^{2}) \mbox{Tr}\left[{1 \over \sqrt{-D^2 - \l_{0} + m^2}}\right]
\eea
which leads to
\bea
 \langle H \rangle 
 &=& \oh \left\langle \mbox{Tr}{-D^2 \over \sqrt{-D^2 - \l_{0} + m^2}}  - 
 \oh \mbox{Tr}{ \l_{0}-m^{2} \over \sqrt{-D^2 - \l_{0} + m^2}} \right\rangle
\non \\
&=&   \oh \left\langle \mbox{Tr} \sqrt{-D^2 - \l_{0} + m^2}    
+ \oh \mbox{Tr}{ \l_{0}-m^{2} \over \sqrt{-D^2 - \l_{0} + m^2}} \right\rangle  
\non \\   
\eea 
Defining
\beq
           \widetilde{k}^{2}_{n} \equiv  \l_{n} - \l_{0}
\eeq
we finally obtain
\beq
 \langle H \rangle = \oh \left\langle \sum_n \left( \sqrt{\widetilde{k}^{2}_{n} + m^2} 
         +  \oh  {\l_{0}-m^{2} \over \sqrt{\widetilde{k}^{2}_{n} + m^2}} \right) \right\rangle
\label{Hnab}
\eeq 

\begin{figure}[t!]
\centerline{\scalebox{0.7}{\includegraphics{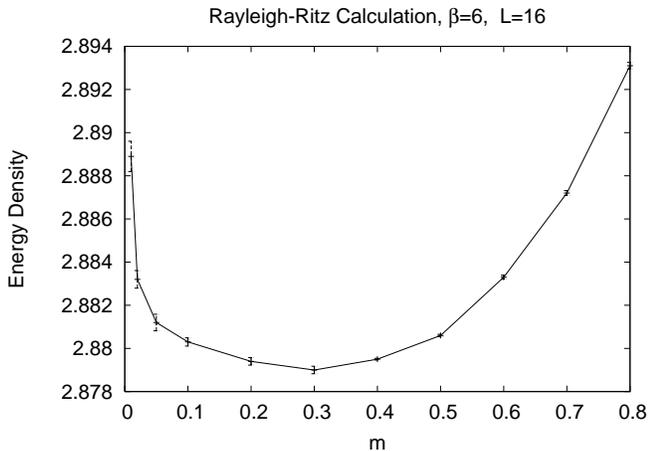}}}
\caption{Vacuum energy $\langle H  \rangle$ of eq.\ \rf{Hnab}, per lattice site,
computed at a variety of mass parameters $m$ on a
$16\times 16$ lattice at lattice coupling $\b=6$.} 
\label{ritz}
\end{figure}

    Suppose that the expectation value of the eigenvalues $\l_{n}$ were independent
of $m^{2}$, with zero variance, as in the free theory.  Setting $\pa \langle H \rangle /\pa m^{2} = 0$, the 
minimum vacuum energy is obtained trivially, at $m^{2}=\langle \l_{0} \rangle$.  In the abelian
free-field limit we have $\l_{0}=0$, so $m=0$ at the minimum and the theory is not confining.  In the
non-abelian theory, in contrast, $\l_{0}>0$, so $m^{2}=\l_{0}>0$ at the minimum, and confinement is obtained.    
Of course, this simple result neglects both the $m^{2}$-dependence of the eigenvalue spectrum,  as
well as contributions arising from functional derivatives of the kernel.  The situation can be improved
on somewhat, at least regarding the $m^2$ dependence, by a numerical treatment. 

    A Monte Carlo evaluation of the energy density $\langle H \rangle/L^{2}$ as a function of $m$,  
for $\b=6$ and $L=16$ and $\langle H \rangle$ as given in eq.\ \rf{Hnab},  is shown 
in Fig.\ \ref{ritz}.  The minimum is away from zero, at roughly $m=0.3$. This gives a string tension which is  
a little low; the known string tension of the Euclidean theory at $\b=6$ would require 
$m=0.515$.  This quantitative disagreement should not be taken too seriously, because the estimate
for vacuum energy on which it is based, eq.\ \rf{Hnab}, is of unknown accuracy.  Once again, in deriving \rf{Hnab},
we have neglected some terms deriving from functional derivatives of the kernel.  Even assuming, as we have,  
that those contributions are quite small (and this has not been shown), they could still have a large effect
on the position of the minimum of a rather flat potential.   The main point of this section is not to obtain $m$
with any degree of accuracy (although that would have been desirable), 
but rather just to see that a non-zero value of $m$,  which implies both confinement and a mass gap, is the
natural outcome of a variational calculation.

\section{\label{sec:KKN} Other Proposals}
 
     There have been other approaches to the Yang-Mills vacuum state in 2+1 dimensions.  In particular,
the vacuum wavefunctional proposed by Karabali, Kim, and Nair (KKN) in ref.\ \cite{KKN} has some 
strong similarities to ours,  and the method we have developed for numerical simulation can be applied
to the KKN vacuum state, as well as to our own proposal.  This application
is important, because we would like to test the claim that a string tension can be derived from the KKN
state which agrees, to within a few percent, with the continuum limit of string tensions extracted 
from lattice Monte Carlo \cite{TepKKN}.

     The KKN approach is formulated in terms of gauge-invariant field variables first introduced by
Bars \cite{Bars}, and the idea is to solve for the ground state of the Hamiltonian, in these variables,
in powers of the inverse coupling $1/g^2$.   To lowest order, when re-expressed in terms of the usual $A$-field
variables, their state has the dimensional reduction form
\beq
           \Psi^{(0)}_0 = \exp\left[- {1\over 4mg^2} \int d^2x ~ B^a(x) B^a(x)] \right]
\label{kkwf1}
\eeq
where
\beq
              m = {g^2 C_A \over 2\pi}
\eeq
and $C_A$ is the quadratic Casimir for the SU(N) group in the adjoint representation.   Because this state
has the dimensional reduction form, the corresponding string tension is easily deduced.   
In lattice units, for the SU(2) group, the predicted string tension is
\beq
             \s_{KKN}^{(0)} = {6 \over \pi \b^2}
\label{s0}
\eeq
which is in rather close agreement with the lattice Monte Carlo results.

    However, the state $ \Psi^{(0)}_0$ is only the first term in a strong-coupling series.  As it stands, it
implies an infinite glueball mass in 2+1 dimensions, and it cannot be even
approximately correct at short distance scales. The question
is whether inclusion of the higher-order terms in the series, which are necessary in order to have
a non-zero correlation length, will affect the long-distance structure, and move the 
prediction for the string tension away from the desired value.  KKN resum all of the terms in the
strong-coupling series which are bilinear in their field variables, and
when this expression is converted back to ordinary $A$-field variables, their resummed
vacuum state has the form
\bea
            \Psi_0 &\approx& \exp\left[- {1\over 2g^2} \int d^2x  d^2y ~ B^a(x) \right.
\non \\
     & &   \qquad  \left. \left( {1 \over \sqrt{-\nabla^2 + m^2} + m } \right)_{xy} B^a(y) \right]
\label{kkwf2}
\eea
This state is gauge non-invariant as it stands, and for that reason must be incomplete.  However, KKN argue that 
the further terms in the strong-coupling series, involving higher powers of the field variables and their 
derivatives, supply the extra terms required to convert the $\nabla^2$ operator in eq.\ \rf{kkwf2} to a covariant
Laplacian.  So, according to ref.\ \cite{KKN}, the vacuum state when re-expressed in ordinary variables
has the form
\bea
            \Psi_0 &\approx& \exp\left[- {1\over 2g^2} \int d^2x  d^2y ~ B^a(x) \right.
\non \\
        & & \qquad  \left. \left( {1 \over \sqrt{-D^2 + m^2} + m } \right)^{ab}_{xy} B^b(y) \right]
\label{kkwf3}
\eea
In this form, the KKN vacuum state is amenable to the numerical methods described above.

        At this point we see that there may be trouble ahead for the previous string tension
prediction.   The problem is that the coefficient $1/(4mg^2)$ in the 
dimensional reduction form \rf{kkwf1} is only obtained if
the lowest eigenvalue $\l_0$ of the covariant Laplacian would be zero.  We know that this
is not the case.  The effective long-distance wavefunctional, gaussian in the $B$-field, 
is obtained as before via a mode cutoff in the $B$-field, and the actual estimate 
for the KKN string tension, according to dimensional reduction, is
\beq
           \s_{KKN} = {3 \over 4 \b} \Bigl\langle  m + \sqrt{\l_0 + m^2}  \Bigr\rangle
\label{skkn}
\eeq
A non-zero $\l_0$ will certainly move the predicted string tensions $\s_{KKN}$ 
away from values given in eq.\ \rf{s0}; the question is by how much.   
This can only be determined, at any given $\b$, by numerical simulation.    

    In Table \ref{table2} we display our results for the string tension $\s_{KKN}$, obtained
by evaluating eq.\ \rf{skkn} in the vacuum state \rf{kkwf3} by the methods developed
in this paper.  It is clear that there is a very substantial discrepancy between the predicted string tension $\s_{KKN}$
and the string tension $\s_{MC}$, obtained by standard Monte Carlo methods in ref.\ \cite{Teper}.  
The disagreement becomes disastrous if $\l_0$ actually diverges, in physical units, in the continuum limit.
In that case the percentage discrepancy at $\b\ra\infty$ will be infinite.  The only way out, 
that we can see, is if eq.\ \rf{kkwf3} is for some reason \emph{not} the true resummation of the 
KKN strong-coupling expansion.

\begin{table}[t!]
\begin{center}
\begin{tabular}{|c|c|c|c|c|c|} \hline
$\beta$ & $L^{2}$      & $\s_{KKN}$     & $\s_{MC}$                        & discrepancy   \\ \hline \hline
     9  &       $24^{2}$  &   0.0340(4)       &  0.0261(2)                 &  30\%            \\ \hline
    12 &       $32^{2}$  &   0.0201(6)       &  0.0139(1)                 &  45\%            \\ \hline
\end{tabular}
\end{center}
\caption{A comparison of the string tension $\s_{KKN}$ calculated numerically from
the Karabali-Kim-Nair vacuum wavefunctional \rf{kkwf3}, by methods developed
above, with the values of the string tension $\s_{MC}$ in $D=3$ dimensions, 
computed by standard lattice Monte Carlo methods in ref.\ \cite{Teper}.}
\label{table2}
\end{table}  
 
         An approach which is closely related to that of KKN, relying on the same
change of field variables, has been followed by Leigh, Minic, and Yelnikov (LMY) in ref.\ \cite{Rob}.
This again results in an expression for the vacuum state which is the exponential
of a bilinear term, with field variables connected by a field-dependent kernel.
Whereas KKN perform a partial resummation of the strong-coupling series to
arrive at their result, LMY rely on a conjectured operator identity
(eq.\ (56) of ref.\ \cite{Rob}) to derive a differential equation for the kernel.  The hope
is that this gives an exact expression for the bilinear term in the wavefunction (of course 
there must be other terms also, because the resulting expression for the 
vacuum is not an exact eigenstate of the Hamiltonian).   Since the derivation relies
on a certain conjecture, the justification for the LMY wavefunctional so far lies
in its predictions.    

     On the one hand, a glueball mass spectrum resulting from the LMY vacuum 
state has been derived, and this spectrum appears to be in very good 
agreement with existing Monte Carlo data.  On the other hand, as in the KKN 
case, the string tension (same as \rf{s0}) and the spectrum are arrived at by 
neglecting the field-dependence of the kernel, which involves a holomorphic-covariant
Laplacian.  We have seen above that neglect of the field dependence of
the kernel can be dangerous, and we think it likely that inclusion
of this field-dependence will affect the LMY string tension and spectrum
significantly.   It may be possible to use the methods developed here to go beyond the 
zero-field approximation for the kernel, as we have done for the KKN state, to get 
a better idea of the true predictions of the LMY state. This is left for future investigation.

\section{\label{sec:nality}The Problem of N-ality}
    
     The Casimir scaling of string tensions is inevitable for the lattice Yang-Mills action in two spacetime
dimensions, and therefore this scaling, out to infinite charged source separations, seems to be 
a consequence of dimensional reduction to two dimensions.  This feature
cannot be true for the asymptotic string tension in 2+1 and 3+1 dimensions, 
except in the $N_{c}=\infty$ limit.  Asymptotic
string tensions in D=2+1 and 3+1 dimensions
must depend only on the N-ality of the charged source, due to color screening
by gluons.   The absence of color screening in $D=2$ dimensions can be attributed to the fact that
a gluon has $D-2$ physical degrees of freedom in $D$ dimensions.  In two dimensions there are no physical degrees
of freedom corresponding to propagating gluons.   If there are no gluons there can be no string-breaking
via dynamical gluons, and hence no N-ality dependence.

      However, the vacuum state of a $d+1$-dimensional gauge theory in temporal gauge does not, in general, describe
a $d$-dimensional Euclidean Yang-Mills theory, despite the fact that each is expressed in terms of a gauge-invariant
combination of $d$-dimensional vector potentials.  For example, the vacuum state of the $2+1$-dimensional 
abelian theory, shown in eq.\ \rf{abelian-vac}, describes the ground state of a theory of free, non-interacting photon
states with a global SU(2) invariance.   Our proposed vacuum state in eq.\ \rf{vacuum} interpolates between a theory of
non-interacting gluons at short distances, and the dimensional reduction form \rf{dimred} at large scales. 
If this is the correct vacuum, then at intermediate distance scales it describes the ground state 
of strongly interacting gluons with physical degrees of freedom; these gluons are free to bind with an external
source.  In that case, the Minkowski-space picture of string-breaking
via gluon pair production should somehow carry over to N-ality dependence for Wilson loops evaluated in the vacuum
state at a fixed time.\footnote{The transition from Casimir scaling to N-ality dependence, due to gluon
string-breaking effects, is very likely to be associated with a vacuum center domain structure,
as discussed recently in ref.\ \cite{g2}. Gluon charge screening and vacuum center domains are simply two
different descriptions, one in terms of particles, the other in terms of fields, of the same effect.}

     At present this is only an optimistic speculation, but the following observation may be relevant:  It is possible
to compute the ground state $\Psi_0[U]$ in strong-coupling Hamiltonian lattice gauge theory, and to identify the
term in that ground state which is responsible for color screening.  From the expansion of this term in powers
of the lattice spacing, we can identify the leading correction to dimensional reduction.  It turns out that
this leading correction has the same form as the leading correction to dimensional reduction that is found in
the proposed vacuum state $\Psi_0[A]$.

\begin{figure}[t!]
\centerline{\scalebox{0.35}{\includegraphics{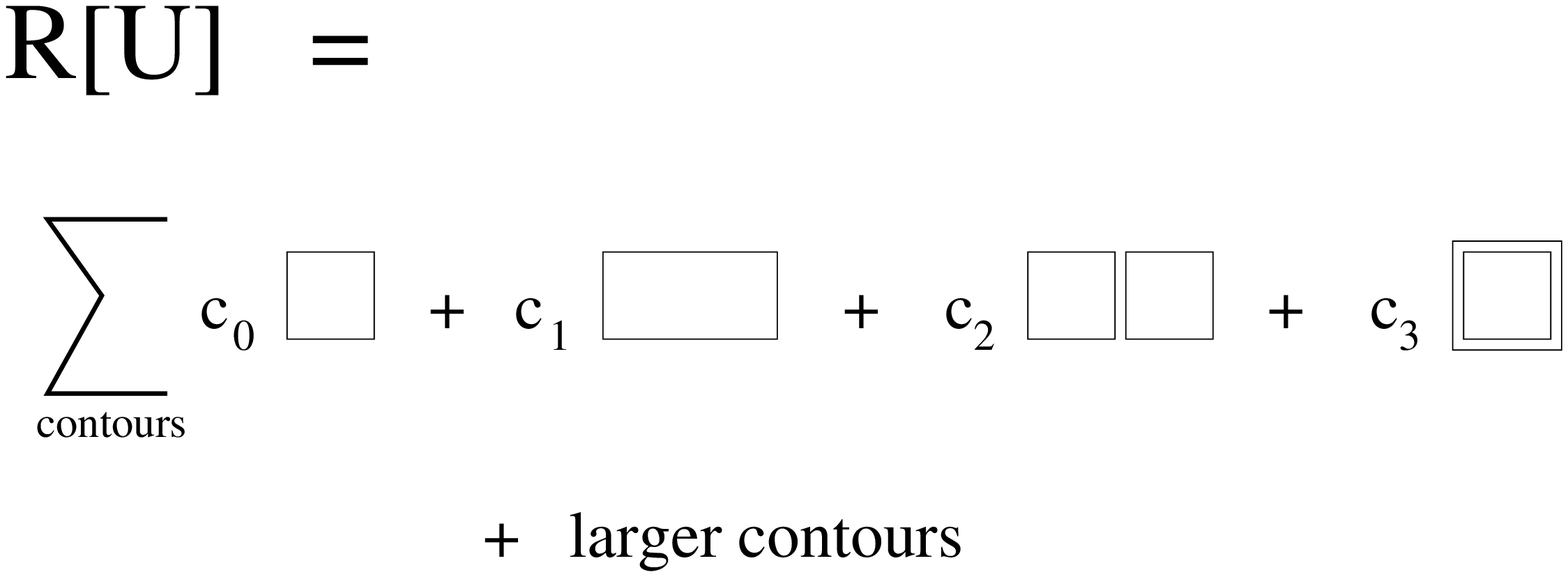}}}
\caption{\label{ru}The first few terms in the strong-coupling expansion of 
the lattice vacuum state $\Psi_0[U]$, with $R[U]=\log(\Psi_0[U])$.} 
\end{figure}

     Denote the lattice vacuum state by $\Psi_0[U] = \exp[R(U)]$.  A strong-coupling technique for calculating $R(U)$
in Hamiltonian lattice gauge theory was developed in ref.\ \cite{Me2}.  In this expansion $R(U)$ is expressed 
as a sum over spacelike Wilson loops and products of loops on the lattice, as indicated schematically in 
Fig.\ \ref{ru}. The coefficient $c_i$ multiplying a contour constructed from (or filled by) $n_P$ 
plaquettes is proportional to $(\b^2)^{n_P}$.  For SU(2) lattice gauge theory in $D=2+1$ dimensions, the first 
few coefficients $c_0,c_1,c_2,c_3$ of the strong-coupling series for $R[U]$ were computed in ref.\ \cite{Guo}.   
The various terms in $R[U]$ can be expanded in a power series in the lattice spacing $a$, and for smoothly 
varying fields it is found that \cite{Guo}
\beq
\Psi_0[U] =
 \exp \left[ - {2\over \b} \int d^2x ~( a \k_0 B^2 - a^3 \k_2 B(-D^2)B + \ldots ) \right] 
\label{scvac}
\eeq
where
\bea
   \k_0 &=& \oh c_0 + 2(c_1+c_2+c_3)
\non \\
   \k_2 &=& \oq c_1 
\eea
and coefficient $c_0$ is $O(\b^2)$, coefficients $c_1,c_2,c_3$ are $O(\b^4)$.

\begin{figure}[t!]
\centerline{\scalebox{0.35}{\includegraphics{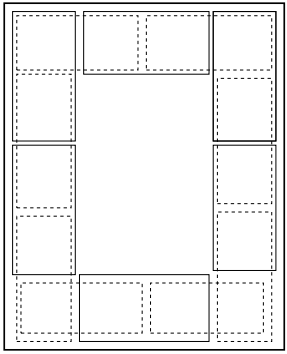}}}
\caption{\label{tile}How $1\times 2$ rectangles in $R[U]$ screen an adjoint
Wilson loop.  The adjoint Wilson loop (in this case with extension $4\times 5$ lattice spacings) 
is denoted by a heavy solid line. The overlapping $1\times 2$ rectangles are indicated 
by (alternately) light solid and light dashed lines. The integration over lattice link variables
yields a finite result, leading to a perimeter-law falloff (eq.\ \rf{perimeter}) 
for large adjoint loops.} 
\end{figure}

   There are several points to note, in connection with eq.\ \rf{scvac}.  First, dimensional reduction
is associated with the term proportional to $\k_0$, which receives contributions from all four terms
shown in Fig.\ \ref{ru}, but the leading correction to dimensional reduction, in the term proportional
to $\k_2$, comes from the $1\times 2$ loop in $R[U]$ proportional to $c_1$.  This is the
contour which couples $B$ (rather than $B^2$) terms in neighboring plaquettes. Secondly, it is not hard to
see that the $1\times 2$ loop in $R[U]$ gives rise to color screening.  Consider evaluating a spacelike
Wilson loop in the adjoint representation
\beq
       W_{adj}[C] = \int DU ~ \mbox{Tr}[U_{adj}(C)] \Psi_0^2[U]
\label{adj}
\eeq
There is a non-zero contribution to the rhs of eq.\ \rf{adj} which comes from lining the perimeter
of the adjoint loop with overlapping $1\times 2$ rectangular loops, as shown in Fig.\ \ref{tile},
deriving from the power series expansion of $\Psi_0^2[U]$.  For a rectangular loop of perimeter
$P(C)$ this diagram gives a perimeter-law contribution
\beq
        \left({c_1 \over 2}\right)^{P(C)-4}
\label{perimeter}
\eeq
to $W_{adj}(C)$.\footnote{Generalizing to an SU(N) theory, it is not hard to show (cf.\  ref.\ \cite{Me2})
that $c_0 \sim 1/g^4 N$, $c_1 \sim 1/g^8 N^3$, and that the perimeter-law contribution shown in Fig.\ \ref{tile}
is down by an overall factor of $1/N^2$
relative to the leading area-law contribution, as it should be.} Thus, the same term that gives the leading 
correction to dimensional reduction is
also responsible for the screening of adjoint loops.  Finally, we note that this leading correction,
proportional to $\k_2$, comes in with a negative sign relative to the $B^2$ term.

   Now let us consider the leading correction to dimensional reduction in the proposed vacuum state
$\Psi_0[A]$ of eq.\ \rf{vacuum}.   The dimensional reduction term was given in eq.\ \rf{dimred1},
and is quadratic in $B^{slow}$.  The definition of $B^{slow}$ in eq.\ \rf{Bslow} involves a mode 
cutoff $n_{max}$, chosen such that $\Delta \l \equiv \l_{n_{max}}-\l_{0} \ll m^{2}$, and 
the first correction to dimensional reduction comes from terms in the vacuum
wavefunctional of order $(\l_n-\l_0)/m^2$, with $n<n_{max}$.  These are obtained from the
$1/m^2$ expansion
\beq
{1 \over \sqrt{-D^2 -\l_0 +m^2}} = {1\over m}\left(1 - {-D^2 -\l_0\over 2m^2} + \ldots \right)
\eeq
Taking the second term in the rhs into account, the part of the vacuum wavefunctional which is
gaussian in $B^{slow}$ is
\beq
     \exp\left[-{1\over m}\int d^2x ~ \left( B^{slow} B^{slow}
             - B^{slow}{-D^2 -\l_0\over 2m^2}B^{slow} + \ldots \right) \right]
\label{leading}
\eeq
where the ellipsis indicates higher powers of the covariant
derivative.  We note the similarity of eq.\ \rf{leading} to the
strong-coupling expression \rf{scvac}.  In particular, there is in
both cases a relative minus sign between the first and second terms.

   The fact that the element responsible for color screening in
$\Psi_0[U]$ generates, in a lattice spacing expansion, the $B(-D^2)B$
term coupling $B$ fields in neighbouring plaquettes, is a hint that it
is this term which might be responsible for the color screening
effect.\footnote{In fact, apart from an overall sign, the $B(-D^2)B$
term looks like the kinetic term of a scalar field in the color
adjoint representation in two Euclidean dimensions.  Matter fields of
that type can, of course, screen adjoint Wilson loops.} If so, the
presence of a very similar correction to dimensional reduction, found in
$\Psi_0[A]$, would presumably give rise to the same effect.

     Of course, it is also possible that the vacuum state \rf{vacuum}
is simply incomplete, and must be supplemented by some additional
terms which are responsible for color screening.  Cornwall
\cite{Cornwall} has recently conjectured that the dimensional
reduction form \rf{dimred} of the vacuum wavefunctional must be altered by
the addition of a gauge-invariant mass term, implemented through the
introduction of a group-valued auxiliary field $\Phi(x)$, i.e.
\beq
  \Psi[A,\Phi] = \exp\left[-\int d^{d}x ~ \{ c_{1}\mbox{Tr}[F_{ij}^{2}] 
         + c_{2} \mbox{Tr}[\Phi^{-1}D_{k}\Phi]^{2}  \} \right]
\eeq
The exponent of this state is stationary around center vortex
solutions, suggesting a vacuum state dominated at large scales by
center vortices.  This would presumbably solve the N-ality problem.
At the moment, however, we lack any direct motivation from the
Schr{\"o}dinger wavefunctional equation for the existence of such a
mass term.  
   
     For a discussion of the N-ality problem in the context of the
KKN approach, see ref.\  \cite{KKN2}.
   
      Another type of contribution which is expected to exist in the
static quark potential is the L{\"u}scher $-\pi (D-2)/24 R$ term.  We
have no insight, at present, as to whether or not this term can be
generated by the proposed vacuum state of eq.\ \rf{vacuum}.

\section{\label{sec:conclusions}Conclusions}

       Our proposal for the ground state of quantized Yang-Mills
theory, in $D=2+1$ dimensions, has a number of virtues.  Apart from
agreeing with the ground state of the free theory in the appropriate
limit, which is a natural starting point for any investigation of this
type, we also find agreement in a highly non-trivial limit, where the
Yang-Mills Schr{\"o}dinger equation is truncated to the zero modes of
the gauge field.  In addition we find, surprisingly, that our vacuum
state is amenable to numerical investigation, despite its very
non-local character.

      We believe that this vacuum state may provide some insight into
the origins of confinement in a non-abelian theory, and the precise
relationship between the mass gap and the string tension.  Confinement
arises here via dimensional reduction, as proposed long ago in ref.\
\cite{Me1}, and this reduction is obtained if the mass parameter $m$
in the vacuum wavefunctional is non-zero.  We have seen that $m\ne 0$
is likely to lower the vacuum energy, in 2+1 dimensions, and this is
related to the fact that in a non-abelian gauge theory the lowest
eigenvalue $\l_{0}$ of the covariant Laplacian is non-zero.  The
relation between $m$ and the asymptotic string tension in 2+1
dimensions is simple, i.e.\ $\s = 3m/4\b$, and if the parameter $m$ is
chosen to produce the string tension known from earlier lattice Monte
Carlo studies \cite{Teper}, then we find that the mass gap extracted
from an appropriate correlator yields a value within 6\% of the mass
gap obtained by standard lattice Monte Carlo methods.

      The most important unresolved question concerns higher
representation string tensions.  At issue is whether corrections to
the simple dimensional reduction limit will convert Casimir scaling to
N-ality dependence, as we have speculated in the previous section, or
whether some additional terms (such as a gauge-invariant mass term
\cite{Cornwall}) are required.  It would also be worthwhile to extend our
considerations to 3+1 dimensions, and to excited-state (glueball and flux-tube) 
wavefunctionals.   These possibilities are currently under investigation.

\acknowledgments{
     We are indebted to Dmitri Diakonov for informing us of his unpublished
results, which motivated the work in section 2 of this article.  We also thank 
Robert Leigh for helpful discussions. This research is supported in part by 
the U.S.\ Department of Energy under 
Grant No.\ DE-FG03-92ER40711 (J.G.), the Slovak Science and Technology 
Assistance Agency under Contract No.\ APVT--51--005704 (\v{S}.O.), 
and the Grant Agency for Science, Project VEGA No.\ 2/6068/2006 (\v{S}.O.).
}

\appendix

\section{\label{sec:appA}3+1 Dimensions}
 
    Although in this article we are mainly interested in the 2+1 dimensional
case, it is worth pointing out that the discussion in section \ref{sec:zmode} can be extended to
3+1 dimensions.   Define
\bea
    S_{3} &=& A_{1}\cdot A_{1} + A_{2}\cdot A_{2}
                     + A_{3}\cdot A_{3}
\non \\
    C_3 &=& (A_{1}\times A_{2})\cdot (A_{1}\times A_{2})
    + (A_{2}\times A_{3})\cdot (A_{2}\times A_{3})
\non \\
   & & \qquad + (A_{3}\times A_{1})\cdot (A_{3}\times A_{1})
\non \\
    D_{3} &=& \Bigl[A_{1}\cdot (A_{2}\times A_{3})]^{2}
\eea
The zero-mode Yang-Mills Hamiltonian is
\beq
     H = -\oh {1\over V} {\pa^{2} \over \pa A_{k}^{a} \pa A_{k}^{a} }
                 + \oh g^{2} V  C_3
\eeq 
Again we express $\Psi_{0}$ as in eq.\ \rf{WKB}, and try to solve $H\Psi_{0}=E_{0}\Psi_{0}$
to leading order in $V$.  This time, with
\beq
    R_{0} = \oh g {C_3\over \sqrt{S_{3}}}
\label{R0d4}
\eeq
we find
\bea
T_{0} &=& V\left[-{\pa R_{0} \over \pa A_{k}^{a}}{\pa R_{0} \over \pa A_{k}^{a}}
+ g^{2} C_3 \right]
\non \\
&=& 0 + g^{2}V \left( {7C_3^{2} \over 4 S_{3}^{2}} - {3 D_{3}\over S_{3}}  \right)
\eea

    In the large volume limit, the ground-state wavefunction will only be non-negligible in
the region of the ``abelian valley", where the zero-mode components 
$A_{1},A_{2},A_{3}$  are nearly aligned, or anti-aligned, in color space.
For definiteness, take the 
large color component (denoted by upper-case $A$) of the color 3-vectors to all lie in the color 3-direction; i.e.
\beq
      A_{1} = \left[ \begin{array}{c} a_{1}^{1} \cr a_{1}^{2} \cr A_{1}^{3} \end{array} \right] ~,~
      A_{2} = \left[ \begin{array}{c} a_{2}^{1} \cr a_{2}^{2} \cr A_{2}^{3} \end{array} \right] ~,~
      A_{3} = \left[ \begin{array}{c} a_{3}^{1} \cr a_{3}^{2} \cr A_{3}^{3} \end{array} \right]
\label{vecA}
\eeq
and lower-case $a$ denotes the small components.  With $a\ll A$ and $VR_{0}\sim O(1)$ it follows that, in the abelian
valley,
\beq
           a \sim {1\over \sqrt{gAV}}
\label{aA}
\eeq
where $A$ and $a$ denote the magnitudes of the large (color 3-direction) and transverse 
field components, respectively.  Since $C_3^2$ and $D_3$ are both $O(a^4)$,
the non-zero terms contributing to $T_{0}$ in eq.\  \rf{T0} are at most of order $1/V^{2}$ and
can be neglected.  Therefore  $\Psi_{0} = \exp[-VR_{0}]$, with $R_{0}$ as given in eq.\ \rf{R0d4},
solves the zero-mode Yang-Mills Schr{\"o}dinger equation to leading order in $V$, in the abelian valley
region away from the origin ($A_k=0 \Rightarrow S_3 = 0$) of field space.

    The generalization of eq.\ \rf{vacuum} to 3+1 dimensions is
\bea
\Psi_0[A] &=& \exp[-Q]
\non \\
 &=&\exp\left[-\oq\int d^3x d^3y ~ F_{ij}^a(x) \right.
\non \\
  & & \qquad \left. \left({1 \over \sqrt{-D^2 - \l_{0} + m^{2}} }
\right)^{ab}_{xy} F_{ij}^b(y) \right]      
\label{vac4d}
\eea 
Again we consider a corner of configuration space in which only the non-zero modes 
make a significant contribution to the wavefunctional, and $|gA|^{2} \gg m_{0}^{2}$.
Then 
\beq
(-D^{2})^{ab}_{xy} = g^{2} \d^{2}(x-y) M^{ab} 
\eeq
as before, with
\beq
            M^{ab} = S_{3}\d^{ab} -  A_{k}^{a}A_{k}^{b}        
\eeq
For a configuration in  the abelian valley, with large components in the color 3 direction
as shown in eq.\ \rf{vecA}, we find
\bea
               Q &=& \oq g V (A_{i}\times A_{j})^{a} (M^{-1/2})^{ab} (A_{i}\times A_{j})^{b}
\non \\
         &\approx& \oq gV (A_{i}\times A_{j})^{a} \left( {\d^{ab}\over \sqrt{S_{3}} }
 - {\d^{a3}\d^{b3}\over \sqrt{S_{3}}}  \right.
\non \\
 & & \qquad \left.  + {\d^{a3}\d^{b3}\over m} \right)
                (A_{i}\times A_{j})^{b} 
\eea
Neglecting the overall coupling and volume factors, the relative orders of magnitude of each of the three 
contributions to $Q$ are as follows:
\bea
          \k_{1} &=&{(A_{i}\times A_{j})\cdot (A_{i}\times A_{j})  
              \over \sqrt{S_{3}} } \sim A a^{2}
\non \\
          \k_{2} &=& {(A_{i}\times A_{j})^{3} (A_{i}\times A_{j})^{3}  
              \over \sqrt{S_{3}} } \sim {a^{4}\over A}
\non \\
           \k_{3} &=&{(A_{i}\times A_{j})^{3} (A_{i}\times A_{j})^{3}  
              \over m} \sim {a^{4}\over m}
\eea
Assume that $\k_{2},\k_{3} \ll \k_{1}$. Then we would have
\beq
          Q =  \oh gV{C_3\over \sqrt{S_{3}}}
\label{Q}
\eeq
and $\Psi_{0}[A]$, evaluated for large zero-mode gauge field
configurations, would agree with the ground
state solution of the zero-mode Yang-Mills Schr{\"o}dinger equation in $D=3+1$ 
dimensions, at least in the neighborhood of the abelian valley.  But we have already
seen that for the solution of the zero-mode equation, the magnitude $a$ of the small 
components is related to the magnitude $A$ of the large components according to \rf{aA}. 
From this it follows that the assumption $\k_{2,3}\ll \k_{1}$ in the abelian valley 
is self-consistent, and justified at large $V$ for $m\ne 0$.

\section{\label{sec:appB}The Spiral Gauge}

Since $\Psi_{0}[A]$ in temporal gauge and 2+1 dimensions 
is gauge-invariant under two-dimensional gauge-transformations, then it is
legitimate to carry out a further gauge-fixing in the two-dimensional plane
when evaluating expectation values
\beq
\langle \Psi_{0}| Q |\Psi_{0} \rangle = \int DA ~ Q[A] \Psi_{0}^{2}
\eeq
In particular, with a complete axial gauge fixing, it is possible to change
variables from $A$ to field-strength $B$ without introducing any further
constraints or field-dependent Jacobian factors, i.e.
\beq
           DA_{1} DA_{2} \ra \mbox{const} \times DB
\eeq
In higher dimensions, as Halpern has shown \cite{Marty}, this change of variables
would be accompanied by a delta function enforcing the Bianchi constraints,
but in two dimensions these constraints are absent.

\begin{figure}[t!]
\centerline{\scalebox{0.45}{\includegraphics{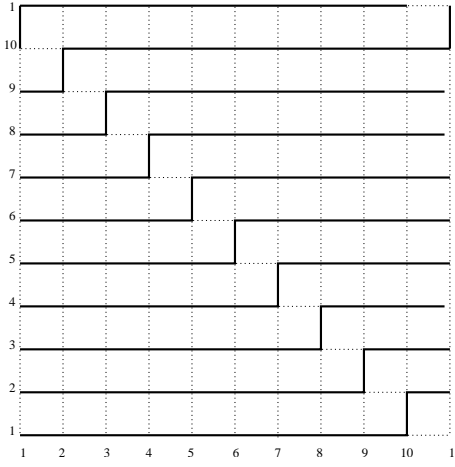}}}
\caption{\label{spiral}The spiral gauge.  Link variables on the solid lines
are set equal to the identity.} 
\end{figure}

    The simplest approach is to set $A_{1}(n_{1},n_{2})=0$ everywhere, where $(n_{1},n_{2})$ are
lattice site coordinates, and invert the
discretized version of $B^{a} = \pa_{1} A_{2}^{a}$ to determine $A_{2}$ from $B$.
The problem with this is that setting $A_{1}=0$ everywhere on a finite, periodic lattice is more than
a gauge choice.  Gauge transformations cannot, in general, set the $A$ field to zero everywhere on
a closed loop, and lines parallel to the $x$-axis are closed by periodicity.  Thus $A_{1}=0$ everywhere
is a boundary condition, as well as a gauge choice.  Although boundary conditions should be
unimportant at sufficiently large lattice volumes, we would still like to keep such artificial conditions
to a minimum, while retaining the simplicity of inverting $B^{a} = \pa_{1} A_{2}^{a}$.  A compromise
is what we will call the ``spiral gauge", in which we set $A=0$ (or link variables $U=I_{2}$) along all links
in a spiral around the toroidal lattice.  An example, on a $10\times 10$ lattice, is shown in Fig.\ \ref{spiral}.
Along the straight sections of the spiral, parallel to the $x$-axis, we have
\beq
       A_2^a(n_1+1,n_2) = B^a(n_1,n_2) + A_2^a(n_1,n_2)
\eeq
For the bent sections, its slightly different.  Referring, e.g., to the bent section in Fig. \ref{spiral}
starting at $n_{1}=9,~n_{2}=1$, we have
\bea
        A_1^a(9,2) &=& - B^a(9,1) - A_2^a(9,1)
\non \\
        A_2^a(10,2) &=&  B^a(9,2) - A_1^a(9,2)
\eea
Now suppose we start out with setting $A_2^a(1,1)=0$.  Applying the above rules all around the spiral we
can get all of the non-zero A-field variables from the B-field variables, but in order to come back to where
we started, with $A_2(1,1)=0$, we have to require that
\beq
        \sum_{n_1,n_2} B^a(n_1,n_2) = 0
\label{sumB}
\eeq
To enforce this condition, we first generate the $B$-field without constraint,
compute the sum
\beq
        S^a = \sum_{n_1,n_2} B^a(n_1,n_2)
\eeq
and then make the readustment
\beq
        B^a(n_1,n_2) \ra B^a(n_1,n_2) - {S^a \over L^2}
\eeq

    So we have done two things beyond just fixing the gauge.  First, the
$A$-field has been set to zero on a single closed spiral around the toroidal
lattice.  Secondly, by setting in addition $A_{2}(1,1)=0$, we have              
imposed a restriction that the $B$-field on the lattice averages to zero
in any given configuration.  These conditions have been imposed for
calculational simplicity; they are not as drastic as setting $A_{1}=0$ on all
links (which sets all Polyakov lines in the $x$-direction equal to unity), and 
ought to be harmless at sufficiently large lattice volumes.

\end{document}